\documentclass[seceq,preprint]{ptptex}

\usepackage{amssymb,graphicx,color,bm,amsmath}
\usepackage{bm}

\usepackage{amsmath}
\usepackage{amssymb}

\newcommand{\wt}{\widetilde}

\newcommand{\ol}{\overline}

\newcommand{\ra}{\rightarrow}

\newcommand{\nn}{\nonumber}

\newcommand{\AD}[1]{$\ol{\mbox{D~\,}}\!\!\!$#1}
\newcommand{\AO}[1]{$\ol{\mbox{O~\,}}\!\!\!$#1}

\newcommand{\tO}[1]{$\wt{\mbox{O~\,}}\!\!\!$#1}

\newcommand{\cN}{{\mathcal N}}

\newcommand{\cO}{{\mathcal O}}
\newcommand{\cA}{{\mathcal A}}

\newcommand{\cG}{{\mathcal G}}

\newcommand{\bR}{\mathbb{R}}

\newcommand{\bZ}{\mathbb{Z}}

\newcommand{\Mkk}{M_{\rm KK}}
\newcommand{\Ukk}{U_{\rm KK}}

\newcommand{\Dint}{D_{\rm rel}}

\def\mat#1{\matt[#1]}
\def\matt[#1,#2,#3,#4]{\left(%
\begin{array}{cc} #1 & #2 \\ #3 & #4 \end{array} \right)}

\def\drawbox#1#2{\hrule height#2pt
        \hbox{\vrule width#2pt height#1pt \kern#1pt
              \vrule width#2pt}
              \hrule height#2pt}

\def\Fund#1#2{\vcenter{\vbox{\drawbox{#1}{#2}}}}
\def\Asymm#1#2{\vcenter{\vbox{\drawbox{#1}{#2}
              \kern-#2pt       
              \drawbox{#1}{#2}}}}

\def\fnd{\Fund{5.5}{0.4}}
\def\asym{\Asymm{5.5}{0.4}}
\def\sym{\fnd\kern-0.4pt\fnd}

\notypesetlogo                       
\preprintnumber[3cm]{
IPMU 09-0085}

\title{
$O(N_c)$ and $USp(N_c)$ QCD from String Theory
}

\author{
Toshiya \textsc{Imoto},$^1$\footnote{E-mail: 
{\tt imoto@eken.phys.nagoya-u.ac.jp}}
Tadakatsu \textsc{Sakai},$^1$\footnote{E-mail: 
{\tt tsakai@eken.phys.nagoya-u.ac.jp}}
and Shigeki \textsc{Sugimoto},$^2$\footnote{E-mail: 
{\tt shigeki.sugimoto@ipmu.jp}}
}

\inst{$^{1}$Department of Physics, Nagoya University,\\
 Nagoya 464-8602, Japan \\ 
$^{2}$Institute for the Physics and Mathematics of the Universe,\\
The University of Tokyo, Chiba 277-8568, Japan}

\abst{
We propose a holographic dual of large $N_c$ quantum chromodynamics
(QCD) with the gauge groups $O(N_c)$ and $USp(N_c)$
and $N_f$ flavors of massless quarks.
This is constructed by adding O6-planes to an intersecting D4-D8 system
in type IIA superstring theory. 
The holographic dual description is formulated in Witten's D4-brane
background with D8-branes and O6-planes embedded in it as probes.
The D4-brane background gives rise to a smooth interpolation of D8-\AD8 pairs
and an O6-\AO6 pair.
We show that the resultant brane configuration explains geometrically
the flavor symmetry breaking patterns in $O(N_c)$ and $USp(N_c)$ 
QCD, which are caused by quark bilinear condensates.
We next discuss that baryons can be realized as D4-\AD4 pairs
wrapped on $S^4$, which intersect with the O6-plane.
By analyzing the tachyons on it, we reproduce the stability conditions
of the baryons that are expected from the gauge theory viewpoint.
The stable baryon configurations are classified systematically 
using K-theory.
We also give a similar analysis of the flux tubes and again
reproduce the results that are consistent with QCD.
}

\begin{document}

\maketitle



\section{Introduction}

Recently, there has been remarkable progress in the study of
the strong coupling dynamics of gauge theories by employing duality
between gauge theory and string theory.
{}Following the discovery of the AdS/CFT correspondence
 \cite{Maldacena,GKP,Witten:AdS} (for a review, see Ref.~\citen{adscft}),
intensive attempts to apply the idea of the gauge/string duality to
QCD have been made.
In Ref.~\citen{SS1}, a holographic dual of QCD with massless quarks is
proposed on the basis of a D4-D8 configuration in type IIA
superstring theory. One of the most insightful results of this model
is that it gives a simple geometric explanation of the chiral symmetry
breaking $U(N_f)_L\times U(N_f)_R\to U(N_f)_V$.

{}From the gauge theory viewpoint, many other interesting models
that are considered to exhibit spontaneous symmetry breaking 
via strong gauge dynamics are known.
Among them is massless QCD with the gauge
group $\cG=O(N_c)$ and $\cG=USp(N_c)$.\footnote{Here
$USp(N)$ is the unitary symplectic group whose element is
$g\in U(N)$ satisfying $g J g^{T}=J$, where $J$ is the anti-symmetric
matrix defined in (\ref{gamma}).
Here, we implicitly assume $N$ to be an even number.
Note that $USp(2)=SU(2)$ by definition.}
The fundamental degrees of freedom of this theory are the gluon and
massless quarks $q_\alpha^{ai}$, where $\alpha=1,2$ denotes the
undotted spinor index, $a=1,2,\cdots,N_c$ is the color index of the gauge
group, and $i=1,2,\cdots,N_f$ is the flavor index.
Since the fundamental representation for these gauge groups is
vector-like, there is no distinction between left-handed and right-handed
components of the quark fields, and hence, the flavor symmetry is $U(N_f)$
for both $\cG=O(N_c)$ and $\cG=USp(N_c)$ cases.\footnote{Here, we include
the anomalous $U(1)$ part of the flavor symmetry, since the effect of
the anomaly vanishes in the large $N_c$ limit. The effect of the anomaly
can be incorporated as studied in \S 5.8 of Ref.~\citen{SS1}.}
Note that $N_f$ must be even for the $\cG=USp(N_c)$ case to avoid
the global anomaly \cite{witten:su2}.

Like $SU(N_c)$ QCD, these models are considered to
develop non-vanishing condensates: 
\begin{align}
\big\langle \epsilon^{\alpha\beta}\,\delta_{ab}
\,q_{\alpha}^{ai}\,q_{\beta}^{bj}\big\rangle&=c\,\delta^{ij} \ ,~~~
(\mbox{for }\cG=O(N_c)) \nn\\
\big\langle \epsilon^{\alpha\beta}\,J_{ab}
\,q_{\alpha}^{ai}\,q_{\beta}^{bj}\big\rangle&=c\,J^{ij} \ .~~~
(\mbox{for }\cG=USp(N_c)) 
\label{condensate}
\end{align}
Here, $c$ is a non-vanishing constant and $J$ is the $USp$ invariant
anti-symmetric tensor. This implies that the flavor symmetry $U(N_f)$ is
spontaneously broken as
\begin{align}
& U(N_f)\to O(N_f) \ ,~~~~~~(\mbox{for }\cG=O(N_c)) \nn\\
& U(N_f)\to USp(N_f) \ .~~~(\mbox{for }\cG=USp(N_c)) 
\label{brpattern}
\end{align}
This phenomenon should result from strong coupling
gauge dynamics, and hence, it is rather difficult to prove that this
really occurs by making a full analysis of the gauge dynamics.
A natural question to ask then is whether we can demonstrate this 
phenomenon from string theory, following the same line in Ref.~\citen{SS1}.

The purpose of this paper is to show that
it is indeed possible to analyze $O(N_c)$ and $USp(N_c)$ QCD
using string theory.
To this end, we begin by constructing a brane configuration
that realizes $O(N_c)$ and $USp(N_c)$ QCD with massless flavors,
by adding O6-planes in the D4-D8 configuration presented in Ref.~\citen{SS1}.
Then, the D4-branes are replaced with the corresponding supergravity
(SUGRA) background obtained in Ref.~\citen{Witten:D4},
which is a holographic dual of large $N_c$ strongly coupled Yang-Mills (YM) theory.
As in Ref.~\citen{SS1}, the quarks are incorporated into the model
by embedding probe D8-branes into the D4-brane background.
Here, we use the probe approximation and ignore the backreaction
of the O6-plane and D8-branes, which can be justified when $N_f\ll N_c$.
It is found that a brane interpolation mechanism that is observed in
Ref.~\citen{SS1} leads to a natural
explanation of the symmetry breaking pattern (\ref{brpattern})
from the physics of intersecting D-branes and O-planes.

We also study the stability of baryons and flux tubes
in $O(N_c)$ and $USp(N_c)$ QCD using our holographic description.
The properties of these objects are quite different from the cases
with $\cG=SU(N_c)$.
As discussed in Ref.~\citen{Witten:ousp}, the baryon number in  $O(N_c)$ QCD
is $\bZ_2$-valued; that means a single baryon is stable, while
two baryons can decay. The flux tubes also behave in a similar way.
On the other hand, the baryons and flux tubes are totally
unstable in $USp(N_c)$ QCD.
In the holographic description of QCD, the baryons and flux tubes
can be realized as D-branes wrapped on $S^4$ in the D4-brane background.
We show that the stability conditions of these D-branes
are in agreement with what we expect in QCD.
Furthermore, since the stable D-brane configurations are classified
using K-theory \cite{Witten:K}, the baryons and flux tubes
correspond to the elements of K-groups. We find that
the relevant K-groups
that are used to 
classify the stable baryons and flux tubes again reproduce
the above results. This refines the classification
via homotopy groups given in Ref.~\citen{Witten:ousp},
in which there are some discrepancies for $N_f\le 3$.

This paper is organized as follows.
In \S 2, we present a brane configuration that defines
massless QCD with the gauge group $\cG=O(N_c)$ and $\cG=USp(N_c)$.
In \S 3, we formulate the holographic dual description
and show that it nicely explains the flavor symmetry breaking
and the stability of baryons and flux tubes in QCD.
We end this paper with a summary and discussion in \S 4.
In Appendices \ref{DO} and \ref{K},
we summarize the properties of intersecting D$p$-O$p'$
systems and K-groups that are used in this paper,
respectively.

\section{Brane configuration of $O(N_c)$ and $USp(N_c)$ QCD}

In this section, we construct a brane configuration of 
$O(N_c)$ and $USp(N_c)$ QCD with massless flavors, by generalizing
the model given in Ref.~\citen{SS1}, which is 
proposed as a holographic dual of $U(N_c)$
QCD with massless flavors.
{}For this purpose, let us first review some key results in Ref.~\citen{SS1}
with an emphasis on how the gluon and quarks emerge.
This model is composed of $N_c$ D4-branes and $N_f$ 
pairs of D8- and \AD8-branes:
\begin{eqnarray}
\begin{array}{c|cccccccccc}
& x^0 & x^1 & x^2 & x^3 & (x^4) & x^5 & x^6 & x^7 & x^8 & x^9 \\
\hline
\mbox{D4} & \circ & \circ & \circ & \circ & \circ &&&&& \\
\mbox{D8-\AD8}
& \circ & \circ & \circ & \circ &  & \circ & \circ & \circ &\circ & \circ 
\end{array}
\label{D4D8}
\end{eqnarray}
Here, the $x^4$ direction is compactified on $S^1$ of radius
$\Mkk^{-1}$. In this paper, we work in the $\Mkk=1$ unit.
The D4-branes are wrapped around this circle, while the
D8-branes and \AD8-branes are located at the antipodal points 
$x^4=\pi/2$ and $x^4=-\pi/2$, 
respectively.
Following Ref.~\citen{Witten:D4}, we impose the anti-periodic
boundary condition along the $S^1$ parametrized by $x^4$
on all the fermions in the system, while all the bosonic
fields are kept periodic.
Then, the gluinos as well as the scalar fields on the D4-brane,
which belong to the adjoint representation of the $SU(N_c)$ gauge
symmetry, become massive.
In addition, the left-handed and right-handed components
of the quark fields ($q_{L}$ and $q_{R}$, respectively)
are created as the massless
modes in the 4-8 strings (open strings stretching from the
D4-branes to the D8-branes) and 4-$\ol 8$ strings, respectively.
Then, the D4-brane world-volume theory flows to four-dimensional
$U(N_c)$ QCD with $N_f$ massless quarks at low energy.\footnote{
We regard the diagonal $U(1)$ part of the $U(N_c)$ gauge symmetry
on the D4-brane as a global symmetry, since it decouples in the IR limit.
}
Note that the gauge symmetries on the D8-branes and \AD8-branes
correspond to the chiral symmetries $U(N_f)_L$ and $U(N_f)_R$,
respectively.

To obtain $O(N_c)$ and $USp(N_c)$ QCD,
we consider an orientifold defined on the basis of the $\bZ_2$ action
$(x^4,x^8,x^9)\ra (-x^4,-x^8,-x^9)$ together with the world-sheet
parity transformation. There are two $(6+1)$-dimensional
fixed planes, called O6-plane,
located at $x^4=x^8=x^9=0$ and $x^4=\pi,\,x^8=x^9=0$,
which are invariant under the $\bZ_2$ action.
To be consistent with the anti-periodic boundary
condition for the fermions around the $S^1$ parametrized by $x^4$, 
the two O6-planes should break different halves of
the supersymmetry \cite{AnDuSa,KaKuSi}.
Whenever we need to distinguish the two, we call the fixed planes
at  $x^4=x^8=x^9=0$ and $x^4=\pi,\,x^8=x^9=0$ O6-plane and
\AO6-plane, respectively.
An \AO6-plane carries an RR charge opposite to that of an O6-plane
with their world-volume orientations opposite to each other.
There are two basic choices of the orientifold $p$-planes called O$p^+$-
and O$p^-$-planes. One way to characterize the O$p^\pm$-planes is to consider
the world-volume gauge theory of probe D-branes, as described
in Appendix \ref{DO}. (See Refs.~\citen{notes,tasi} for reviews.)

\begin{figure}[http]
\begin{center}
\includegraphics[scale=0.80]{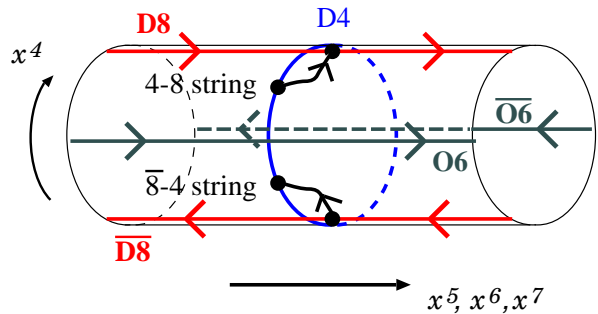}
\parbox{75ex}{\caption{\small{
D4-D8-O6 configuration.
}}
\label{ddo}
}
\end{center}
\end{figure}

Now, we argue that massless QCD with $\cG=O(N_c)$ and $\cG=USp(N_c)$
can be obtained by placing O6$^+$-\AO6$^+$
and O6$^-$-\AO6$^-$ into (\ref{D4D8}), respectively. The brane
configuration we consider is
\begin{eqnarray}
\begin{array}{c|cccccccccc}
& x^0 & x^1 & x^2 & x^3 & (x^4) & x^5 & x^6 & x^7 & x^8 & x^9 \\
\hline
\mbox{D4} & \circ & \circ & \circ & \circ & \circ &&&&& \\
\mbox{D8-\AD8}
& \circ & \circ & \circ & \circ &  & \circ & \circ & \circ &\circ & \circ 
\\
\mbox{O6$^{\pm}$-\AO6$^{\pm}$} & \circ & \circ & \circ & \circ
& & \circ& \circ& \circ & 
\end{array}
\label{D4D8O6}
\end{eqnarray}
(See also Fig.~\ref{ddo}).
In general,
 a system with a D$p$-brane and an O$p'$-plane
that are extended along $(q+1)$ common directions
is characterized by the number of relative transverse directions
$\Dint\equiv p+p'-2q$. For our D4-O6 and D8-O6 configurations
in (\ref{D4D8O6}), we have $\Dint=4$ and $\Dint=2$, respectively.
Then, as explained in Appendix \ref{DO}, the $U(N_c)$ gauge field on the
D4-branes obeys
\begin{eqnarray}
 A_\mu(x^\mu,x^4)=-\gamma_\pm  A_\mu^T(x^\mu,-x^4) \gamma_\pm^{-1}
\label{O6D4}
\end{eqnarray}
($\mu=0,\dots,3$) for the cases with O6$^\pm$-planes,
together with the periodic boundary condition
$A_\mu(x^\mu,x^4+2\pi)=A_\mu(x^\mu,x^4)$.
Here, $\gamma_\pm$ is defined by
\begin{eqnarray}
\gamma_+=I_{N_c}\ ,~~\gamma_-=J_{N_c}\equiv \mat{0,I_{N_c/2},-I_{N_c/2},0}
\label{gamma}
\end{eqnarray}
with $I_k$ being the identity matrix of rank $k$. Note that $N_c$ should
be even for the case with O6$^-$-plane.
Therefore, the zero mode of the gauge field along $x^4$
that survives the orientifold projection (\ref{O6D4}) gives four-dimensional
$O(N_c)$ and $USp(N_c)$ gauge fields for the cases with O6$^+$- and
O6$^-$-planes, respectively.
On the other hand, the orientifold action
($\bZ_2$ action associated with the orientifold)
maps D8-branes to \AD8-branes and vice versa. This implies that
the quark fields are given by $\bZ_2$-invariant linear 
combinations of $q_L$ and $q_R^*$, because the $\bZ_2$ maps
a 4-8 string to an $\ol 8$-4 string, and the flavor symmetry
is reduced to $U(N_f)$. 
Therefore, we conclude that the brane configuration (\ref{D4D8O6})
with O6$^+$-plane and  O6$^-$-plane
yields $O(N_c)$ and $USp(N_c)$ QCD with $N_f$ massless
quarks, respectively.\footnote{
It is also possible to construct $O(N_c)$ and $USp(N_c)$ QCD by
adding orientifolds in the D4-D6 system considered in Ref.~\citen{KrMaMyWi}.
However, the full flavor symmetry $U(N_f)$ is not
manifestly realized in such brane configurations.
}

A few comments are in order.
The number of Weyl fermions in
the fundamental representation of $USp(N_c)$ gauge group must
be even to avoid the global anomaly \cite{witten:su2}.
This fact can be understood in the brane setup (\ref{D4D8O6}) as follows.
Recall that a D8-brane is a source of the RR nine-form potential and its
field strength is dual to RR zero-form field strength $F_0$, which takes
an integer value if we normalize $F_0=1$ for a unit flux.
In (\ref{D4D8O6}), the ten-dimensional space-time is divided 
by the D8- and \AD8-branes into two regions,
 $-\pi/2<x^4<\pi/2$ and $\pi/2<x^4< 3\pi/2$. The difference in the
$F_0$ flux between the two regions is $N_f$. Let us consider what
happens if $N_f$ is odd. Without loss of generality,
we assume that $F_0$ is even in the region $\pi/2<x^4< 3\pi/2$.
Then, the O6$^-$-plane at $x^4=0$ is embedded in the region
with odd $F_0$ flux. However, it is known that the O6$^-$-plane is allowed
to exist only in the background with even $F_0$ flux \cite{HyImSu}.
This is exactly what we expect to avoid the global anomaly.
Instead of the O6$^-$-plane, one could consider an \tO6$^-$-plane,
which is an O6$^-$-plane with a D6-brane stuck on it
and allowed to exist 
in the background with odd $F_0$ flux \cite{HyImSu}.
In this case, we have an additional flavor of quarks created
by the open string stretched between the \tO6$^-$-plane and
the D4-branes, and the number of flavors
will again become even.
In this way, the number of flavors is always
even and the global anomaly is automatically avoided.

\section{Analysis in holographic dual of QCD}

In this section, we study the properties of $O(N_c)$ and 
$USp(N_c)$ QCD with massless flavors by working in the
holographic dual of the brane configuration given in
(\ref{D4D8O6}).

\subsection{Flavor symmetry breaking}
\label{FSB}
We start by discussing how the flavor symmetry breaking
patterns of $O(N_c)$ and $USp(N_c)$ QCD in (\ref{brpattern})
can be seen in the holographic dual description obtained from
the brane setup (\ref{D4D8O6}).
{}For this purpose, we first provide a brief review of
holographic chiral symmetry breaking of $U(N_c)$ QCD
\cite{SS1}.

In general gauge/string duality,
the holographic dual description is obtained by replacing the
D-branes representing the gauge theory with the corresponding curved
background that solves the SUGRA equations of motion. In our brane
configuration (\ref{D4D8O6}), the SUGRA background corresponding to the
$N_c$ D4-branes is given by Witten
\cite{Witten:D4}. Although the explicit solution is known, we only need
information on the topology of the background, which is
$\bR^{1,3}\times\bR^2\times S^4$. Here, $\bR^{1,3}$ is the
four-dimensional Minkowski space-time $\{x^{0\sim 3}\}$ and $\bR^2$ is
the two-dimensional cigar-like geometry parametrized by $x^4$ and the
radial coordinate $r$ of the five-dimensional plane $\{x^{5\sim 9}\}$
transverse to the D4-branes.\footnote{
Here, this $r$ is related to $\sqrt{(U/\Ukk)^3-1}$ in the coordinates
used in Ref.~\citen{SS1}.
}
 We also use the coordinates $(y,z)$ related to $(r,x^4)$ as
\begin{eqnarray}
 y=r\cos x^4\ ,~~~ z=r\sin x^4\ .
\end{eqnarray}
The $S^4$ corresponds to the angular directions in the five-dimensional
plane $\{x^{5\sim 9}\}$,
around which $N_c$ units of an RR four-form field strength are turned on.
This flux plays a substantial role in analyses of baryons to be
made in \S \ref{baryon}.

Assuming $N_c\gg N_f$, we use the probe approximation to include
the D8-branes. Namely, the D8-branes are embedded in Witten's
D4-brane background as probes and their backreaction is neglected.
This probe approximation is related to the quenched approximation,
which is widely used in lattice QCD, since quark loops in QCD correspond
to holes in the string world-sheet attached to the D8-branes.
Then, we find that the D8-branes and
\AD8-branes in the previous brane configuration (\ref{D4D8})
must be smoothly connected when we replace the D4-brane with the
corresponding SUGRA background described above and we end up with
the configuration with only one connected component of $N_f$ D8-branes
extended along the $x^{0\sim 3}$, $z$, and $S^4$ directions.
It is shown in Ref.~\citen{SS1} that this
smooth interpolation of the D8-\AD8 pairs
gives the geometric realization of the chiral symmetry breaking.
Since D8-branes and \AD8-branes are located at $x^4=\pi/2$ and
$x^4=-\pi/2$ in the previous brane configuration (\ref{D4D8}),
the chiral symmetries $U(N_f)_L$ and $U(N_f)_R$ correspond
to the gauge symmetry at the two boundaries $z\ra+\infty$ and
 $z\ra-\infty$, respectively.\footnote{
The gauge symmetry at the boundaries $z\ra\pm\infty$ yields the
global symmetry in QCD, since the corresponding gauge couplings vanish.
(See Ref.~\citen{SS2}.)
}
Now, the $z$-dependent gauge transformation shifts the $z$ component
of the gauge field $A_z$ on the D8-brane world-volume, and hence,
it is spontaneously broken. (See \S \ref{pion} for more on this
point.) The unbroken part is given by the constant gauge transformation
that is $U(N_f)_V$. Therefore, the spontaneous chiral symmetry
breaking $U(N_f)_L\times U(N_f)_R\ra U(N_f)_V$ is caused by the fact
that the geometry of the D4-brane background requires that
the D8-branes and \AD8-branes be smoothly connected.

\begin{figure}[http]
\begin{center}
\includegraphics[scale=0.80]{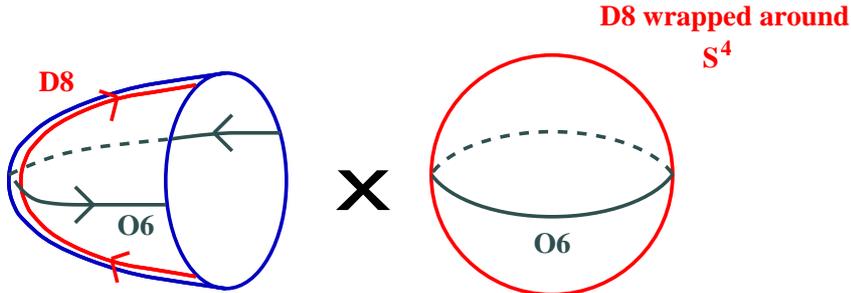}
\parbox{75ex}{\caption{\small{
D8-O6 system in Witten's D4 background.
The cigar-like geometry on the left side is parametrized by $(y,z)$,
and the D8-branes and O6-plane are extended along the $z$ and $y$
 directions, respectively. The D8-branes are wrapped around the $S^4$
on the right side, while the O6-plane is wrapped on an $S^2$
embedded in the $S^4$ via $x^8=x^9=0$.
}}
\label{cigar}
}
\end{center}
\end{figure}

Next, we argue that the flavor symmetry breaking in 
$O(N_c)$ and $USp(N_c)$ massless QCD can again be
understood geometrically.
As explained in the previous section, we consider the orientifold defined
by the $\bZ_2$ action $(x^4,x^8,x^9)\ra (-x^4,-x^8,-x^9)$. Since the
D4-brane background has this $\bZ_2$ symmetry, we can consistently 
impose the orientifold projection in the holographic dual
description.\footnote{Here, the backreaction of the O6-planes on the
geometry is also neglected by assuming large $N_c$.}
Then, we find that the fixed plane of the $\bZ_2$ orientifold action
is an O6$^\pm$-plane extended along the $x^{0\sim  3}$, $y$, and $S^2$
directions located at $z=x^8=x^9=0$. (See Fig.~\ref{cigar}.)
This means that
the O6$^\pm$-\AO6$^\pm$ pair in the previous configuration
(\ref{D4D8O6})  (see also Fig.~\ref{ddo}) is now smoothly connected in the
D4-brane background,
which is possible because the world-volumes of the O6$^\pm$-plane and  
\AO6$^\pm$-plane in (\ref{D4D8O6}) are oriented opposite to each other,
just as in the case of the D8-\AD8 pairs.
Together with the probe D8-branes considered above,
we obtain an intersecting D8-O6 system with $\Dint=4$.
Then, as explained in Appendix \ref{DO}, the $U(N_f)$ gauge field
on the D8-brane world-volume satisfies
\begin{eqnarray}
 A_\mu(x^\mu,z)=-\gamma_\pm  A_\mu^T(x^\mu,-z) \gamma_\pm^{-1}
\ ,~~
 A_z(x^\mu,z)=\gamma_\pm  A_z^T(x^\mu,-z) \gamma_\pm^{-1}
\label{O6D8}
\end{eqnarray}
($\mu=0,\dots,3$) for the cases with O6$^\pm$-planes, where $\gamma_\pm$
is defined by replacing $N_c$ with $N_f$ in (\ref{gamma}).
Here, we have omitted the components of the gauge field
along the $S^4$ as well as
the $S^4$ coordinate dependence for simplicity.
The gauge transformation consistent with the constraint (\ref{O6D8})
is given by a gauge function $g(x^\mu,z)\in U(N_f)$ satisfying
\begin{eqnarray}
 g(x^\mu,z)= \gamma_\pm g^*(x^\mu,-z) \gamma_\pm^{-1}\ ,
\label{g}
\end{eqnarray}
where $g^*=(g^{-1})^T$ is the complex conjugate of $g$. In particular,
the gauge transformation at $z=0$ is restricted to $O(N_f)$
or $USp(N_f)$ for O6$^+$-plane and O6$^-$-plane, respectively.

With these setups, we are now ready to show the flavor symmetry breaking
patterns in $O(N_c)$ and $USp(N_c)$ QCD.
{}First, the D8-branes have only a single boundary,
because the apparent two boundaries at $z\ra\pm\infty$ are identified with
each other by the orientifold projection, which involves $z\ra -z$.
Thus, the flavor symmetry is $U(N_f)$, as discussed in the previous section.
{}Next, by noting that the unbroken flavor symmetry results from 
constant gauge transformations on the D8-branes, we find
that the unbroken flavor symmetry consistent with the
constraint (\ref{g}) is $O(N_f)$ for $O(N_c)$ QCD (with O6$^+$-plane) and
$USp(N_f)$ for $USp(N_c)$ QCD (with O6$^-$-plane).
This is exactly what we have observed in (\ref{brpattern}).
In the holographic description, the spontaneous flavor symmetry breaking
(\ref{brpattern}) is caused by the fact that the D8-branes must
intersect with the O6$^\pm$-plane in the D4-brane background.

\subsection{Meson effective action}
\label{pion}

As argued in Refs.~\citen{SS1,SS2}, the open strings attached to the probe
D8-branes are interpreted as mesons. The meson effective action is
given by the effective action on the D8-brane world-volume.
After reducing the $S^4$ directions,\footnote{
Here, we only consider the five-dimensional components of the
nine-dimensional gauge field on the D8-brane world-volume
that is constant along the $S^4$.
Then, the constraint (\ref{constraint1}) imposed on the gauge field
is reduced to (\ref{O6D8}).
The higher KK-modes associated with the $S^4$ are the artifacts of the model
that cannot be interpreted as the bound states in QCD.}
the effective theory of the mesons in $U(N_c)$ QCD
turned out to be a five-dimensional $U(N_f)$
 Yang-Mills (YM) - Chern-Simons (CS) theory. The action is
given by Ref.~\citen{SS1}
\begin{align}
&S_{\rm 5 dim}\simeq S_{\rm YM}+S_{\rm CS}\ ,\nn\\
&S_{\rm YM}=-\kappa\int d^4 xdz\,
{\rm Tr}\left(\frac{1}{2}h(z)F_{\mu\nu}^2+k(z)F_{\mu z}^2\right)\ ,
~~S_{\rm CS}=\frac{N_c}{24\pi^2}\int_{\rm 5 dim}\omega_5(A)\ ,
\label{5dim}
\end{align}
where $\mu,\nu=0,\cdots,3$ are the Lorentz indices
for the four-dimensional space-time and $z$ is the coordinate
of the fifth dimension along the D8-brane world-volume.
The warp factors in the YM action are given by $k(z)=1+z^2$
 and $h(z)=(1+z^2)^{-1/3}$,
and $\omega_5(A)$ is the Chern-Simons five-form.
The Kaluza-Klein decomposition of the five-dimensional gauge field
with respect to the warped $z$-direction yields
an infinite tower of massive vector/axial-vector mesons as well as
a massless pion
in four-dimensional space-time. It was shown
in Refs.~\citen{SS1,SS2} that the resultant four-dimensional effective
action for these mesons reproduces various properties of the mesons
found in the experiments.

The low energy effective theory of mesons for $O(N_c)$ and
$USp(N_c)$ QCD can be obtained by simply imposing the constraint
(\ref{O6D8}) on the five-dimensional gauge field in (\ref{5dim})
and restrict the $z$ integral to $0\le z<+\infty$.
In fact, the $\bZ_2$ action associated with the 
O6$^{+}$-plane and O6$^{-}$-plane
 corresponds to the charge conjugation
and G-parity transformation, respectively, and it is easy to check that
the action (\ref{5dim}) is invariant under them.

It is interesting to show how to extract the pion degrees of
freedom that appear as the Nambu-Goldstone modes associated with the
spontaneous flavor symmetry breaking (\ref{brpattern}).
Following Ref.~\citen{SS1}, let us define $U(N_f)$-valued
fields
\begin{eqnarray}
\xi_\pm^{-1}(x^\mu)={\rm P}\exp\left(
-\int_0^{\pm\infty} dz \, A_z(x^\mu,z)\right)\ .
\end{eqnarray}
Here,  $\xi_+(x^\mu)$ and $\xi_-(x^\mu)$ are related to
each other by the constraint (\ref{O6D8}) as
\begin{eqnarray}
\xi_+(x^\mu) = \gamma_\pm\xi_-^*(x^\mu)\gamma_\pm^{-1}\ .
\end{eqnarray}
These fields transform as
\begin{eqnarray}
\xi_\pm(x^\mu)\ra h(x^\mu)\xi_\pm(x^\mu) g_\pm^{-1}
\label{xi}
\end{eqnarray}
under the five-dimensional gauge transformation given by the
gauge function satisfying constraint (\ref{g}), where\footnote{
Here, we assume $g_\pm$ to be constant, since the gauge symmetry at
$z\ra\pm\infty$ corresponds to the global symmetry in QCD. It is
also useful to gauge it by allowing the $x^\mu$-dependence to $g_\pm$
in order to extract the current associated with the
flavor symmetry. (See Refs.~\citen{SS1,SS2,HSS1}.)}
\begin{eqnarray}
 h(x^\mu)=g(x^\mu,0)\ ,~~ g_\pm=\lim_{z\ra\pm\infty} g(x^\mu,z)\ .
\end{eqnarray}
As discussed in \S \ref{FSB}, the gauge transformation at the
boundary $g_\pm\in G$ corresponds to the flavor symmetry $G=U(N_f)$ of
the $O(N_c)$ and $USp(N_c)$ QCD. On the other hand,
$h(x^\mu)$ corresponds to the hidden local symmetry introduced
in Ref.~\citen{BaKuUeYaYa} to write down an effective action including
pion and rho mesons. In our case, the constraint (\ref{g}) implies
$h(x^\mu)\in H$ with $H=O(N_f)$ and $H=USp(N_f)$
for $O(N_c)$ and $USp(N_c)$ QCD, respectively.
The transformation property (\ref{xi}) is exactly
the same as that for the Nambu-Goldstone mode associated with
the spontaneous symmetry breaking $G\ra H$ in the general formulation
of the hidden local symmetry approach. (See Ref.~\citen{BKY} for a review.)
Since $H$ here is a local symmetry, the physical Nambu-Goldstone mode
parametrizes the quotient space $G/H$.

\subsection{Baryons and flux tubes}
\label{baryon}

As in the $\cG=SU(N_c)$ case, a baryon in QCD with $\cG=O(N_c)$ and $\cG=USp(N_c)$ 
is composed of $N_c$ quarks with the
color indices totally anti-symmetrized by the $\epsilon$-tensor.
It is argued, however, that the baryons
exhibit novel features regarding the stability \cite{Witten:ousp}.
For $\cG=O(N_c)$, the conserved baryon number is $\bZ_2$-valued.
Namely, a single baryon is stable, while two baryons can decay.
This is because the product
of two $\epsilon$-tensors can be decomposed as 
\begin{align}
\epsilon_{a_1\cdots a_{N_c}}\,\epsilon_{b_1\cdots b_{N_c}}
= \delta_{a_1b_1}\,\delta_{a_2b_2}\cdots \delta_{a_{N_c}b_{N_c}}
\pm \cdots \ ,
\label{eeddd}
\end{align}
implying that two baryons can decay into $N_c$ mesons.
{}In contrast, none of the $\cG=USp(N_c)$ baryons is stable, because
the $\epsilon$-tensor can be decomposed as
\begin{align}
 \epsilon_{a_1\cdots a_{N_c}}=J_{a_1a_2}J_{a_3a_4}\cdots 
J_{a_{N_c/2-1}\,a_{N_c/2}} \pm\cdots \ .
\label{ejjj}
\end{align}
This implies that even a single baryon is unstable against decay to
$N_c/2$ mesons.

Now, we demonstrate that the stability of baryons in $O(N_c)$ and
$USp(N_c)$ QCD can be understood geometrically
from the holographic description formulated in \S \ref{FSB}.
To this end, it is helpful to recall first that
the baryons in QCD with $\cG=U(N_c)$ can be realized 
as a D4-brane wrapped around $S^4$ at the tip of the cigar
($y=z=0$) in the D4-brane background \cite{Witten:baryon,SS1}.
It was shown in Ref.~\citen{Witten:baryon}
that $N_c$ units of electric charge are induced on the wrapped
D4-brane through the RR four-form field strength around $S^4$,
and the Gauss law constraint requires that the induced charge 
be cancelled by $N_c$ open strings ending on the D4-brane.
The resultant configuration is interpreted as a bound state of $N_c$ quarks,
namely, a baryon. $n$-baryon systems can be obtained by wrapping $n$ D4-branes
around $S^4$.
Without the probe D8-branes, the $N_c$ strings attached to a
D4-brane are extended to infinity and the baryon is infinitely massive.
In our brane configuration with the D8-branes,
the fundamental strings can end on D8-branes and the baryon
becomes dynamical.

It would be natural to guess that the baryons in $O(N_c)$ and $USp(N_c)$
QCD can also be obtained by wrapping D4-branes around the $S^4$.
However, it turns out that the baryons are constructed using D4-\AD4 pairs
wrapped around the $S^4$.
A key ingredient here is that
a D4-brane wrapped on $S^4$ intersects with the O6$^\pm$-plane
with $\Dint=6$. Then, the orientifold action 
maps the D4-branes to \AD4-branes, and they should be paired
in order to be invariant under the orientifold action. 
If we consider $n$ D4-\AD4 pairs wrapped on the $S^4$,
the tachyon field $T$ created by the $4$-$\ol 4$ and
$\ol 4$-$4$ strings becomes an $n\times n$ complex matrix satisfying
\begin{eqnarray}
T(x^8,x^9)=\mp T^T(-x^8,-x^9)
\label{constT}
\end{eqnarray}
for the case with O6$^{\pm}$-plane, respectively. (See (\ref{constraint3}).)
For the case with O6$^-$-plane, the tachyon field is allowed to take
a constant non-vanishing value for any $n$.
Thus, we conclude that the baryons in $USp(N_c)$ QCD are unstable,
as expected in the field theoretical consideration reviewed above.
Let us next consider the case with O6$^{+}$-plane
that corresponds to the $O(N_c)$ QCD. When $n=1$,
the constraint (\ref{constT}) implies that the tachyon field vanishes
at $x^8=x^9=0$ and the D4-\AD4 pair cannot be annihilated completely.
A tachyon profile consistent with (\ref{constT}) for $n=1$ is given by
a $k$-vortex configuration $T(x^8,x^9)\propto (x^8+ix^9)^k$ with odd $k$,
which gives rise to the decay of the D4-\AD4 pair
to $k$ D2-branes wrapped around $S^2$ in $S^4$ at $x^8=x^9=0$.\footnote{
See, for example, Ref.~\citen{Sen} for a review of tachyon condensation
in unstable D-brane systems.}
Note that these D2-branes are embedded in the O6$^+$-plane with $\Dint=4$.
It is found from (\ref{constraint1}) that the gauge symmetry on
the D2-brane world-volume is $O(k)$, and the
scalar fields on it corresponding to the position of the D2-branes
transverse to the O6$^+$-plane are in the anti-symmetric representation of
the gauge group $O(k)$.
For $k=1$, because there is no anti-symmetric representation in this case, 
a single D2-brane cannot move away from the O6$^+$-plane and
gives a stable configuration. However,
an even number of D2-branes can be separated from the O6$^+$-plane and
they will eventually decay, since the two-cycles in $S^4$ are
contractible.
After all, for odd $k$, we end up with
a single D2-brane stuck on the O6$^+$-plane, which
is a stable configuration interpreted as a baryon.
When $n=2$, a constant non-vanishing tachyon field
is consistent with the constraint (\ref{constT}) and
the D4-\AD4 pairs can be annihilated completely.
Therefore, we see that a single baryon in $O(N_c)$ QCD is stable,
while two baryons can decay. This is exactly what is expected
from the field theoretical argument.

To examine the nature of the baryon vertex in more detail,
we study the effects of open strings attached on a D4-\AD4 pair wrapped on
the $S^4$ for the $\cG=O(N_c)$ case. As in the $\cG=U(N_c)$ case reviewed above,
there exist $N_c$ open strings\footnote{
The number of strings here is counted in the
covering space before the orientifold projection.
}
attached to the D4-brane wrapped on the $S^4$, and the other end
points of these $N_c$ open strings can be attached to the probe
D8-branes. These 4-8 strings are mapped to 8-$\ol 4$ strings
under the orientifold action.
Then, this configuration is interpreted as a color singlet bound state of
$N_c$ quarks.
Note that a 4-8 string and an 8-$\ol 4$ string
that are {\it not} mirror to each other under the orientifold action
can be recombined to form a 4-$\ol 4$ string and an 8-8 string.\footnote{
Here, if the 4-8 and 8-$\ol 4$ strings were
mirror to each other, the midpoint of the resultant 4-$\ol 4$ string 
would be confined at the O6$^+$-plane, which is not allowed unless
there are D-branes stuck on the O6-plane.
}
Here, the 8-8 string is interpreted as a meson and it can move away from
the baryon, and the 4-$\ol 4$ string is no longer carrying the flavor indices.
This is interpreted as the process that
two quarks in the baryon are transmuted to a gluon through the gauge
interaction.
Unlike the $U(N_c)$ QCD, the number of quarks trapped in a baryon is
not conserved, since there is no $U(1)$ symmetry
associated with the quark number charge conserved in a
baryon.\footnote{
In this paper, we use the terminology ``baryon''
for a color singlet bound state of quarks and/or gluons
with $N_c$ color indices contracted by an epsilon tensor.
The $\bZ_2$-valued baryon number in the $O(N_c)$ QCD can be defined
by the $\bZ_2$ generated by the element of $O(N_c)/SO(N_c)$.}
Repeating this process, we can construct a bound state
of $k$ gluons and $(N_c-2k)$ quarks
\begin{eqnarray}
 \epsilon_{a_1a_2\cdots a_{N_c-1}a_{N_c}}
F^{a_1a_2}_{\mu_1\mu_2}\cdots
F^{a_{2k-1}a_{2k}}_{\mu_{2k-1}\mu_{2k}}q^{a_{2k+1}i_{2k+1}}
\cdots q^{a_{N_c}i_{N_c}}\ ,
\label{FFqq}
\end{eqnarray}
where $F^{ab}_{\mu\nu}$ is the
field strength of the $O(N_c)$ gauge field and $q^{a i}$
is the quark field. This state
corresponds to a D4-\AD4 pair with $2k$ 4-$\ol 4$ strings
and $(N_c-2k)$ 4-8 and 8-$\ol 4$ strings attached to it
in the covering space of the orientifold.

For even $N_c$, we can take $k=N_c/2$ in (\ref{FFqq})
and construct a bound state given by the Pfaffian of
the gluon field. This Pfaffian particle exists even for 
pure Yang-Mills theory without quarks, and can be 
constructed in an alternative manner parallel to that
shown in Ref.~\citen{Witten:baryon},
in which a holographic dual of $\cN=4$ super Yang-Mills theory
with $\cG=O(N_c)$ and $\cG=USp(N_c)$ is constructed by considering
D3-O3 systems. To see this, we note that
pure Yang-Mills theory with gauge group $\cG=O(N_c)$ and $\cG=USp(N_c)$
can also be formulated by placing an O4$^\mp$-plane parallel to the
D4-branes instead of the O6$^\pm$-\AO6$^\pm$ pair
in (\ref{D4D8O6}) without adding D8-\AD8 pairs.
Then, the holographic dual of this system is given by Witten's
D4 background with the $S^4$ divided by the $\bZ_2$ orientifold action,
which maps a point in the $S^4$ to the antipodal point.
The Pfaffian particle in this setup is given by a D4-\AD4 pair
wrapped around the $S^4$, which will decay to a stable D2-brane
wrapping a $\bZ_2$ invariant $S^2$ in the $S^4$ for the case with
O4$^-$-plane.

When $N_c$ is odd, 
$(N_c-1)$ pairs of 4-8 and 8-$\ol 4$
strings can be converted to $(N_c-1)$ 4-$\ol 4$ strings through the
process considered above, but at least a pair of 4-8 and 8-$\ol 4$
strings that are mirror to each other under the orientifold action
remains connected to the D8-brane.
This is plausible because 
the baryons are fermionic, forbidding
all the quarks in the
baryons to be replaced with gluons.
This fact leads us to an interesting consequence.
As discussed above, the tachyon condensation with a one-vortex
profile gives rise to the decay of the D4-\AD4 pair to a D2-brane 
wrapped on the $S^2$
along the O6$^+$-plane inside the $S^4$. The above consideration
implies that for odd $N_c$,
there must be an odd number of 2-8 strings (together with the 8-2
strings that are their images of the orientifold action)
attached on the D2-brane.
It would be interesting to see if this fact can be understood directly
from the D2-brane world-volume theory.

The stability of the baryons can also be understood from the 
topological viewpoint.
A simple argument for it was given in Ref.~\citen{Witten:ousp} 
assuming that the baryons are
described as a topological soliton in the effective theory of 
Nambu-Goldstone boson associated with
the spontaneous flavor symmetry breaking as in the Skyrme model \cite{Skyrme}.
The Nambu-Goldstone mode takes a value in the coset space $G/H$,
where $G=U(N_f)$ is the flavor symmetry group and $H$ is the unbroken
subgroup, that is, $H=O(N_f)$ and $H=USp(N_f)$ for $O(N_c)$ and
$USp(N_c)$ QCD, respectively.
Then, the classification of the point-like 
topological solitons in $(3+1)$-dimensions is given by
the homotopy group
\begin{align}
& \pi_3(U(N_f)/O(N_f))=\bZ_2 \ ,~~~
(\mbox{for }\cG=O(N_c)\mbox{ with }N_f\ge 4 ) \nn\\
& \pi_3(U(N_f)/USp(N_f))=0 \ .~~
(\mbox{for }\cG=USp(N_c))
\label{pi3}
\end{align}
This is in accord with the previous results on the stability of baryons.
However, for the cases with less than four flavors, we have
$\pi_3(U(N_f)/O(N_f))=\bZ_4$ for $N_f=3$ and $\pi_3(U(N_f)/O(N_f))=\bZ$
for $N_f=2$, which do not agree with what we expect in $O(N_c)$ QCD.
It was argued in Ref.~\citen{Witten:ousp} that this discrepancy might be due
to the fact that the topological classification in (\ref{pi3}) only
takes into account the Nambu-Goldstone boson, which is only
a part of the whole configuration space including an infinite tower of
massive mesons.

This subtlety can be improved by using
our holographic description. We first recall that the stable baryons
are described as stable D-brane configurations.
It is now well established that the topological classification of the
D-branes should be made using K-theory rather than the homotopy
group \cite{Witten:K}.
As shown in Appendix \ref{K}, the topological classification
of the baryons is given by the following K-groups:
\begin{align}
& KR(\bR^{1,3})=\bZ_2 \ ,~~(\mbox{for O6}^{+}) \nn\\
& KH(\bR^{1,3})=0 \ .~~~~(\mbox{for O6}^{-}) 
\label{Kbaryon}
\end{align}
Here, $\bR^{p,q}$ denotes $\bR^{p+q}$ with an involution acting on
$\bR^p$. In (\ref{Kbaryon}), $\bR^{1,3}$ corresponds to the
four-dimensional space $\{(z,x^1,x^2,x^3)\}$ transverse to the baryon.
Roughly speaking, the K-groups (\ref{Kbaryon}) classify the
vector bundles associated with the D8-branes with an involution
consistent with the constraint (\ref{O6D8}). (See Appendix \ref{K} for
more details.)
Our result (\ref{Kbaryon}) is consistent with the classification of
stable baryon configurations found above.
It is known that the K-groups (\ref{Kbaryon}) are isomorphic to the
homotopy groups (\ref{pi3}) for large $N_f$. The main difference
between the homotopy group and K-theory is that the K-theory allows
the creation and annihilation of D-brane - anti-D-brane pairs.
A field configuration that looks stable in terms of the
homotopy group may decay through the creation and annihilation of
D-brane - anti-D-brane pairs, and our observation suggests that
it actually happens in $O(N_c)$ QCD with $N_f<4$.

There is another type of gauge-invariant configurations in QCD, i.e.,
flux tubes. Consider a pair of heavy quarks that belong to a
representation $R$ and its complex conjugate $\ol R$ of the gauge group
$\cG$. Without the dynamical quarks, the electric flux sourced by the heavy
quarks is squeezed to a straight line to form a flux tube. This is
responsible for the linear potential between the quarks.
The situation is changed when the fundamental quarks are added.
For $\cG=SU(N_c)$ and $\cG=USp(N_c)$, any heavy quark is screened by the
pair-created fundamental quarks because any representation $R$ can be
constructed from tensor products of an appropriate number of 
fundamental and anti-fundamental representations.
Hence, any flux tube is unstable for $SU(N_c)$ and $USp(N_c)$ QCD.
In contrast, flux tubes in $O(N_c)$ QCD exhibit
a $\bZ_2$ stability, because
the flux tube composed of a heavy quark pair in the spinor
representation cannot be screened by the fundamental quarks, and
therefore, is stable, while the tensor product of two spinors can
be screened \cite{Witten:ousp}.

The stability of the flux tube in $O(N_c)$ and $USp(N_c)$ QCD
can again be understood from our holographic description.
The obvious string-like object in the holographic description
is the fundamental string. However, this corresponds to the
flux tube created by quarks in the fundamental representation
of the gauge group, which is expected to be unstable. In fact,
the fundamental strings can break up with the end points on the
D8-brane, and are sent to infinity on the D8-brane world-volume.
A possible candidate for the stable flux tube in the holographic
description is a non-BPS D5-brane wrapped on the $S^4$.
Since it is a six-dimensional object wrapped on a four-cycle,
it also behaves as a string in the four-dimensional world.
The number of relative transverse directions of the non-BPS D5-brane
and O6$^\pm$-plane is $\Dint=5$.
The tachyon field $T$ on $n$ non-BPS D5-branes is an $n\times n$ Hermitian
matrix satisfying 
\begin{eqnarray}
\wt T(x^8,x^9)=\mp \wt T^T(-x^8,-x^9)\ ,
\label{constT2}
\end{eqnarray}
where $\wt T=T\gamma_\pm$,
for the case with O6$^{\pm}$-plane, respectively.
(See (\ref{constraint6}).)
As in the discussion for baryons above, the tachyon field can take a
constant non-vanishing value for the case with O6$^-$-plane, which 
results in the annihilation of the non-BPS D5-brane.
In fact, by using
the equation of motion for the gauge field on the non-BPS D5-brane
\begin{eqnarray}
d*F\simeq c'\, F_4\wedge e^{-T^2} dT \ ,
\label{nonD5}
\end{eqnarray}
with $F_4$ being the RR four-form field strength in the background
and $c'$ a non-vanishing constant,
it turns out that the electric flux along the
flux tube is induced through the process of tachyon condensation.
Here, we only consider the diagonal $U(1)$ part of the gauge field and
tachyon field, and we have used the CS term for the non-BPS D-brane
obtained in Refs.~\citen{KrLa,TaTeUe}.
Then, because the electric flux is interpreted as fundamental strings,
the non-BPS D5-brane is transmuted into fundamental strings,
and hence, it is unstable in the presence of the D8-brane
as discussed above.
This is consistent with the fact that the flux tube is unstable for
$USp(N_c)$ QCD.
For the case with O6$^+$-plane, we see that an even number of non-BPS
D5-branes are unstable for the same reason.
When $n=1$, the tachyon
field is a real scalar field, and a profile consistent with the
constraint (\ref{constT2}) is given by a kink configuration
$T(x^8,x^9)\propto x^8$. Then, the non-BPS D5-brane decays to a D4-brane
wrapped on the $S^3\subset S^4$ at $x^8=0$. Similar to the D2-brane
representing a baryon discussed above, this D4-brane is hooked at the
O6$^+$-plane with $\Dint=4$ and gives a stable configuration.
Therefore, we conclude that a single flux tube is stable, while two flux
tubes can decay, which is again consistent with the $\bZ_2$ stability
observed in the $O(N_c)$ QCD.

It is also interesting to note that the tension of the flux tube
created by a heavy quark pair in the spinor representation in the
$O(N_c)$ QCD is of order $N_c$. This fact is consistent with the above
interpretation, because the tension of D-branes is proportional to
$1/g_s\simeq\cO(N_c)$
\cite{Witten:ousp,Witten:baryon}.

It was argued in Ref.~\citen{Witten:ousp} that the flux tubes can also be
understood as topological solitons in the effective theory of
the Nambu-Goldstone mode. Then, a topological classification
is given by
\begin{align}
&\pi_2(U(N_f)/O(N_f))=\bZ_2 \ ,~~~
(\mbox{for }\cG=O(N_c)\mbox{ with }N_f\ge 3 ) \nn\\
&\pi_2(U(N_f)/USp(N_f))=0 \ .~~
(\mbox{for }\cG=USp(N_c)) 
\end{align}
Again, this argument can be improved using K-theory.
As discussed in Appendix \ref{K}, the relevant K-groups for the flux tubes are
\begin{align}
& KR(\bR^{1,2})=\bZ_2 \ ,~~(\mbox{for O6}^{+}) \nn\\
& KH(\bR^{1,2})=0 \ .~~~~(\mbox{for O6}^{-}) 
\end{align}
This is in accord with the classification of stable flux tubes given
above.

It is also possible to construct a domain wall by wrapping
a D6-brane around the $S^4$ \cite{Witten:theta}.
However, this domain wall is not stable
in the presence of the probe D8-brane.
The D6-brane can be represented as a magnetic flux
in the D8-brane world-volume gauge theory.
It can be shown that the magnetic flux is
smeared in an infinite volume and the domain
wall becomes infinitely thick. Then, the magnetic
flux per unit volume will vanish and become invisible.

\section{Summary and discussion}

In this paper, we have studied the strong coupling dynamics of
large $N_c$ $O(N_c)$ and $USp(N_c)$ QCD with $N_f$ massless
quarks using string theory.
We formulated QCD with $\cG=O(N_c)$ and $\cG=USp(N_c)$
using the intersecting D4-D8-O6 
system (\ref{D4D8O6}), and the holographic dual description was
given by the
intersecting D8-O6 system in Witten's D4-brane background
(Fig.~\ref{cigar}), which is 
valid for $N_f/N_c \ll 1$ and large 't Hooft coupling.

With this machinery, we first showed that the flavor symmetry $U(N_f)$
is broken in a manner consistent with the gauge theory viewpoint
(\ref{brpattern}).
Although we have not given a direct proof of a non-vanishing quark 
bilinear condensate, our result can be regarded as a strong evidence of
the spontaneous flavor symmetry breaking.
Next, we discussed the stability of baryons and flux tubes.
The baryons and flux tubes can be realized as D4-\AD4 pairs
and non-BPS D5-branes wrapped on $S^4$ in the string dual picture,
respectively.
By examining the tachyon modes on them,
we derived exactly the same stability conditions
as those expected from the gauge theory viewpoint.
The classification of the stable configurations is
systematically given using K-theory, which improves
the subtlety of the classification by homotopy groups.

One of the interesting features of the gauge/string duality is that
information on the gauge group is translated to the geometry
in the string theory description.
As we have seen in (\ref{condensate}), (\ref{eeddd}), and (\ref{ejjj}),
the properties of the gauge-invariant quark bilinear operators
as well as the stability of the baryons and flux tubes in QCD
follow from the structure of the gauge groups in addition
to the strong dynamics of QCD.
On the other hand, in the holographic dual description,
the D-branes responsible for the color indices are
replaced with the SUGRA background and all information
on the gauge-invariant operators in the gauge theory
is encoded in the geometry of the background.
As we have seen, the flavor symmetry breaking and the stability of
baryons and flux tubes are all understood from the topology of the
background.

These successes lead us to expect that the
gauge/string duality is sufficiently powerful to explore strong coupling
dynamics of a wide class of gauge theories including non-supersymmetric
and non-conformal field theories.
A deeper investigation along this line would be worthwhile.

\section*{Acknowledgements}

We would like to thank our colleagues at the Institute
for the Physics and Mathematics of the Universe (IPMU)
and the particle theory group at Nagoya University
for helpful discussions. S.S. is especially grateful to
O. Bergman, K. Hori, K. Saito, and A. Tsuchiya for valuable 
discussions.
Part of the work of T.S. was carried out while he was visiting IPMU.
The work of S.S. is supported  in part by 
the Grant-in-Aid for Young Scientists (B), Ministry of Education, 
Culture, Sports, Science and Technology (MEXT), Japan,
JSPS Grant-in-Aid for Creative Scientific Research
No. 19GS0219 and also by World Premier International
Research Center Initiative (WPI Initiative), MEXT, Japan. 
We are also grateful to the Yukawa Institute for Theoretical Physics
at Kyoto University. Discussions during the YITP workshop YITP-W-09-04 on
``Development of Quantum Field Theory and String Theory'' were useful
to complete this work. 

\appendix

\section{D$p$-O$p'$ system}
\label{DO}

In this appendix, we summarize some important properties of 
intersecting D$p$-O$p'$ systems
that are used in this paper. Here, 
$p'$ is assumed to be even or odd for type IIA or IIB superstring 
theory, respectively.

Before considering the general D$p$-O$p'$ system, let us first recall
the results in the cases of $p'=9$.
The string theory with O9$^-$-plane is known as the type I string theory
and the case with O$9^+$-plane is the string theory
with $USp(32)$ gauge symmetry considered in Refs.~\citen{Su,SchWi}.
The gauge groups, tachyon, and scalar fields on the D$p$-brane
world-volume in the presence of the O$9^\pm$-plane
are summarized in Table \ref{typeI}.
See Refs.~\citen{Bergman:2000tm,AsSuTe,GaHo} for the derivation.
\begin{table}[htb]
\begin{center}
$$
\begin{array}{c|cccl}
\hline\hline
p&\mbox{gauge}&\mbox{tachyon}&\mbox{scalar}\\
\hline
9&USp(N)&\mbox{none}&\mbox{none}\\
8&USp(N)&\sym&\asym\\
7&U(N)&\sym&\mbox{adj.}\\
6&O(N)&\sym&\sym\\
5&O(N)&\mbox{none}&\sym\\
4&O(N)&\asym&\sym\\
3&U(N)&\asym&\mbox{adj.}\\
2&USp(N)&\asym&\asym\\
1&USp(N)&\mbox{none}&\asym\\
0&USp(N)&\sym&\asym\\
-1&U(N)&\sym&\mbox{adj.}\\
\hline
\end{array}
~~~~~~~~~
\begin{array}{c|cccl}
\hline\hline
p&\mbox{gauge}&\mbox{tachyon}&\mbox{scalar}\\
\hline
9&O(N)&\mbox{none}&\mbox{none}\\
8&O(N)&\asym&\sym\\
7&U(N)&\asym&\mbox{adj.}\\
6&USp(N)&\asym&\asym\\
5&USp(N)&\mbox{none}&\asym\\
4&USp(N)&\sym&\asym\\
3&U(N)&\sym&\mbox{adj.}\\
2&O(N)&\sym&\sym\\
1&O(N)&\mbox{none}&\sym\\
0&O(N)&\asym&\sym\\
-1&U(N)&\asym&\mbox{adj.}\\
\hline
\end{array}
$$
\parbox{75ex}{
\caption{\small
Gauge groups, tachyon, and scalar fields on
the world-volume of $N$ D$p$-branes with the O$9^+$-plane (left)
or O$9^-$-plane (right).
}
\label{typeI}
}
\end{center}
\end{table}

Next, consider an intersecting D$p$-O$p'$ system such that the
D$p$-brane and O$p'^{\pm}$-plane are extended along $(q+1)$ common directions.
To be more specific, we consider $N$ D$p$-branes extended along
 $x^0,x^1,\dots,x^p$ directions and
an O$p'^{\pm}$-plane extended along
 $x^0,x^1,\dots,x^q,x^{p+1},\dots,x^{p+p'-q}$ directions.
In this appendix, we use the indices $\mu$, $i$,
$I$, and $M$ for the coordinates $x^\mu=(x^0,\dots,x^q)$, $x^i=(x^{q+1},\dots,x^p)$, 
$x^I=(x^{p+1},\dots,x^{p+p'-q})$, and $x^M=(x^{p+p'-q+1},\dots,x^9)$, respectively.
The $\bZ_2$ orientifold action 
acts on these coordinates as $(x^\mu,x^i,x^I,x^M)\ra (x^\mu,-x^i,x^I,-x^M)$.
\begin{eqnarray}
\begin{array}{c|cccc}
& x^\mu & x^i & x^I & x^M  \\
\hline
\mbox{D$p$} & \circ & \circ & \\
\mbox{O$p'^{\pm}$} & \circ && \circ &
\end{array}
\end{eqnarray}
This system is characterized by the number of relative transverse
directions parametrized by $\{x^i\}$ and $\{x^I\}$:
 $\Dint\equiv (p-q)+(p'-q)=p+p'-2q$.
Note that $\Dint$ is invariant under T-duality.
The D$p$-branes and O$p'^{\pm}$-plane
in the above configuration are mapped to D$(9-\Dint)$-branes
and O9$^\pm$-plane, respectively, by T-dualizing all the directions
transverse to the O$p'^\pm$-plane.

For $\Dint\equiv 0$ (mod 4), the massless bosonic fields on the
D$p$-brane are the gauge
fields $A_\mu$, $A_i$ and scalar fields $\Phi_I$, $\Phi_M$. These fields
belong to the adjoint representation of the $U(N)$ gauge group.
It is useful to define $\cA=(A_\mu,\Phi_M)$ and $\varphi=(A_i,\Phi_I)$,
which are related to the gauge field and scalar fields
on the D$(9-\Dint)$-brane 
in the T-dualized picture, respectively.
The orientifold projection requires that the fields $\cA$ and $\varphi$ at
$(x^\mu,x^i)$ are related to their transpose $\cA^T$ and $\varphi^T$ at
$(x^\mu,-x^i)$, respectively. The explicit relations can be obtained
{}from the consistency with Table \ref{typeI} in the T-dualized picture as
\begin{align}
& \cA(x^\mu,x^i)=-\gamma_\mp \cA^T(x^\mu,-x^i)\gamma_\mp^{-1}\ ,~~
 \varphi(x^\mu,x^i)=+\gamma_\mp \varphi^T(x^\mu,-x^i)\gamma_\mp^{-1}\ ,\nn\\
&\hspace{50ex}~\mbox{(for O$p'^\pm$, $\Dint=0,8$)}\nn\\
& \cA(x^\mu,x^i)=-\gamma_\pm \cA^T(x^\mu,-x^i)\gamma_\pm^{-1}\ ,~~
 \varphi(x^\mu,x^i)=+\gamma_\pm \varphi^T(x^\mu,-x^i)\gamma_\pm^{-1}\ .\nn\\
&\hspace{50ex}~\mbox{(for O$p'^\pm$, $\Dint=4$)}
\label{constraint1}
\end{align}
where $\gamma_+=I_N$ and $\gamma_-=J_N$ are defined in (\ref{gamma}).

For $\Dint\equiv 2$ (mod 4), the orientifold action
maps D$p$-brane to \AD$p$-brane, and hence,
we ought to consider $N$ D$p$-\AD$p$ pairs to obtain a 
$\bZ_2$ invariant brane configuration. Then,
the gauge group of the system becomes $U(N)\times U(N)$,
and the gauge fields and scalar fields are paired as
 $(\cA,\wt\cA)$ and $(\varphi,\wt\varphi)$. In addition,
it is important to consider the tachyon field $T$
created by the open strings stretched between the D$p$-branes and 
\AD$p$-branes, which belongs to the bifundamental representation
of the $U(N)\times U(N)$ gauge group. The constraints for these fields
are
\begin{eqnarray}
 \cA(x^\mu,x^i)=-\wt\cA^T(x^\mu,-x^i)\ ,~~~
 \varphi(x^\mu,x^i)=+\wt\varphi^T(x^\mu,-x^i)
\label{constraint2}
\end{eqnarray}
and
\begin{align}
&T(x^\mu,x^i)=\pm T^T(x^\mu,-x^i)\ ,~~~(\mbox{for O$p'^\pm$,
 $\Dint=2,10$})\nn\\
&T(x^\mu,x^i)=\mp T^T(x^\mu,-x^i)\ .~~~(\mbox{for O$p'^\pm$,
 $\Dint=6$})
\label{constraint3}
\end{align}
Again, these relations are obtained from the consistency
with Table \ref{typeI} in the T-dualized picture.

For odd $\Dint$, the D$p$-brane
is a non-BPS D-brane whose world-volume theory is
a $U(N)$ gauge theory with a tachyon field in the adjoint representation.
The constraints are obtained as
\begin{align}
& \cA(x^\mu,x^i)=-\gamma_\mp \cA^T(x^\mu,-x^i)\gamma_\mp^{-1}\ ,~
 \varphi(x^\mu,x^i)=+\gamma_\mp \varphi^T(x^\mu,-x^i)\gamma_\mp^{-1}\ ,\nn\\
&\hspace{50ex}~(\mbox{for O$p'^\pm$, $\Dint=1,7,9$})\nn\\
& \cA(x^\mu,x^i)=-\gamma_\pm \cA^T(x^\mu,-x^i)\gamma_\pm^{-1}\ ,~
 \varphi(x^\mu,x^i)=+\gamma_\pm \varphi^T(x^\mu,-x^i)\gamma_\pm^{-1}\ ,\nn\\
&\hspace{50ex}~(\mbox{for O$p'^\pm$, $\Dint=3,5$})
\label{constraint4}
\end{align}
\begin{align}
& T(x^\mu,x^i)=-\gamma_\mp T^T(x^\mu,-x^i)\gamma_\mp^{-1}\ ,
~~~(\mbox{for O$p'^\pm$, $\Dint=1,9$})\nn\\
& T(x^\mu,x^i)=+\gamma_\pm T^T(x^\mu,-x^i)\gamma_\pm^{-1}\ ,
~~~(\mbox{for O$p'^\pm$, $\Dint=3$})\nn\\
& T(x^\mu,x^i)=-\gamma_\pm T^T(x^\mu,-x^i)\gamma_\pm^{-1}\ ,
~~~(\mbox{for O$p'^\pm$, $\Dint=5$})\nn\\
& T(x^\mu,x^i)=+\gamma_\mp T^T(x^\mu,-x^i)\gamma_\mp^{-1}\ .
~~~(\mbox{for O$p'^\pm$, $\Dint=7$})
\label{constraint5}
\end{align}
Or, if we define $\wt T\equiv T\gamma_\pm$
(for $\Dint=3,5$) or
 $\wt T\equiv T\gamma_\mp$
(for $\Dint=1,7,9$), the condition
(\ref{constraint5}) can be written as
\begin{align}
& \wt T(x^\mu,x^i)=\pm \wt T^T(x^\mu,-x^i)\ ,
~~~(\mbox{for O$p'^\pm$, $\Dint=1,3,9$})\nn\\
& \wt T(x^\mu,x^i)=\mp \wt T^T(x^\mu,-x^i)\ .
~~~(\mbox{for O$p'^\pm$, $\Dint=5,7$})
\label{constraint6}
\end{align}

\section{Real K-theory}
\label{K}

Here, we summarize some properties of the K-theory, which are used
in the paper.\footnote{S.S. thanks A. Tsuchiya and K. Hori for
the discussion on K-theory, which was crucial for this appendix.}

Let $X$ be a manifold with an involution.
The Real K-theory group $KR(X)$ is defined as the Grothendieck
group of the category of complex vector bundles $E$ over $X$ provided
with an anti-linear involution that commutes with the involution
of $X$. (See Refs.~\citen{Atiyah,Karoubi} for more details.)
The physical interpretation is as follows. Consider a
topological classification of the gauge configurations
on the D-brane - anti-D-brane system filling $X$.
The vector bundle $E$ corresponds to
the Chan-Paton bundle of the D-brane on which the gauge field is defined
as a connection. The involution of $X$ corresponds to the
orientifold action $x^i\ra -x^i$
considered in Appendix \ref{DO} and the gauge field is required to
obey the constraint (\ref{constraint1}) with $\gamma_+$.
$KR(X)$ is the group that classifies the configuration of
this D-brane - anti-D-brane system
allowing the creation and annihilation of
D-brane - anti-D-brane pairs. As proposed in Ref.~\citen{Witten:K},
this is considered to be equivalent to the classification
of D-brane charge in this system. If $X$ is non-compact,
this K-group only takes into account the charge of the D-branes
whose world-volumes are not extended to infinity.
It is also possible to define
a similar group denoted as $KH(X)$ by using $\gamma_-$ in the
constraint (\ref{constraint1}).
 (See Refs.~\citen{Gukov,Hori,BeGiHo,BeGiSu}
 for more details.)

We use the notation $\bR^{p,q}$ that
denotes $\bR^{p+q}=\bR^p\times\bR^q$
with an involution acting on the first $\bR^p$ factor,
and $S^{p,q}$ is the $(p+q-1)$-dimensional unit sphere
in $\bR^{p,q}$. It can be shown that $KR(X\times\bR^{p,q})$
only depends on $p-q$, and we define
\begin{eqnarray}
 KR^{p-q}(X)= KR(X\times\bR^{p,q})\ ,
\label{KRpq}
\end{eqnarray}
and similarly for $KH^{-n}(X)$.
Some useful properties are
\begin{eqnarray}
 KR^{-n}(X)=KR^{-n+8}(X)\ ,~~
 KH^{-n}(X)=KR^{-n+4}(X)\ .
\end{eqnarray}

These groups $KR^{-n}(X)$ and $KH^{-n}(X)$
are used to classify the D-brane charge
in type II string theory with an orientifold in the following way.
Let us consider a
ten-dimensional space-time
of the form $\bR^{r,s}\times \bR^{r',s'}\times X$,
where $r+s+r'+s'+\dim X=10$.
We assume that the set of fixed points of the involution
on $X$ is a connected $k$-dimensional
submanifold of $X$ and consider an O$p'^-$-plane
with $p'+1=s+s'+k$ placed at the fix point of the involution
in the ten-dimensional space-time.
Then, $KR^{-n}(X)$
with $n=9-p'-r'+s'$ is used to classify
the D-brane charge for stable D-branes
whose world-volumes are uniformly extended
along $\bR^{r,s}$ and localized in $\bR^{r',s'}$.
Similarly, $KH^{-n}(X)$ is used for the case with
an O$p'^+$-plane. It is easy to see that
this statement agrees with the proposal given in Ref.~\citen{Hori}
by noting
\begin{align}
&KR^{-n}(X)=KR^{-(9-p')}(\bR^{r',s'}\times X)\ ,~~~
(\mbox{for O$p'^-$}) \nn\\
&KH^{-n}(X)=KH^{-(9-p')}(\bR^{r',s'}\times X)\ ,~~~
(\mbox{for O$p'^+$})
\label{KRKH}
\end{align}
which
follow from the relation (\ref{KRpq}).
Note that if we consider
D$p$-branes whose world-volume is given by
$\bR^{r,s}\times X$, where $r+s+\dim X=p+1$,
the D$p$-O$p'$ system has $\Dint=r+s'+\dim X - k=n$.
The elements of $KR^{-n}(X)$ and $KH^{-n}(X)$ are interpreted
as the charge of stable D-branes
made by the D$p$-branes (or D$p$-\AD$p$ system) whose
world-volume field configuration is uniform
along $\bR^{r,s}$.

To calculate the K-groups relevant to our system,
the following relation will be useful.
\begin{align}
 KR^{-n}(\bR^{p,q}\times S^{r,s})= 
KR^{-n+p-q}(pt) \oplus KR^{-n+p-q+r-s+1}(pt)~~~
(\mbox{for $s\ge 1$})
\label{KKK}
\end{align}
Here $pt$ denotes a point.
 (See below for a derivation.)
Note that (\ref{KKK})
also holds for $s=0$ with $r\ge 3$ \cite{Atiyah}.
The $KR^{-n}(pt)$ is known as
\begin{eqnarray}
\begin{array}{c|ccccccccc}
\hline\hline
 n ~\mbox{(mod 8)}&0&1&2&3&4&5&6&7\\
\hline
 KR^{-n}(pt)&\bZ&\bZ_2&\bZ_2&0&\bZ&0&0&0\\
\hline
\end{array}
\end{eqnarray}
Note that (\ref{KKK}) can be written as
\begin{align}
&KR^{-n}(\bR^{p,q}\times S^{r,s})= 
KR^{-n}(\bR^{p,q}) \oplus KR^{-n}(\bR^{p+r,q+s-1})\ ,\nn\\
&KH^{-n}(\bR^{p,q}\times S^{r,s})= 
KH^{-n}(\bR^{p,q}) \oplus KH^{-n}(\bR^{p+r,q+s-1})\ ,
\end{align}
{}from which we interpret that $KR^{-n}(\bR^{p,q})$
and $KH^{-n}(\bR^{p,q})$ correspond to the D-branes
localized along $\bR^{p,q}$ and
wrapped on the $S^{r,q}$, while
$KR(\bR^{p+r,q+s-1})$ and $KH(\bR^{p+r,q+s-1})$
correspond to the D-branes localized along
$\bR^{p,q}\times S^{r,s}$.

Let us apply these results to the system considered
in \S \ref{FSB}.
We are interested in the space-time
 $\bR^{1,5}\times S^{2,3}$ where
 $\bR^{1,5}=\{(z,y,x^0,\dots,x^3)\}$ and
 $S^{2,3}$ is the unit four-sphere in $\{(x^5,\dots,x^9)\}$
with the involution $(z,x^8,x^9)\ra(-z,-x^8,-x^9)$.  

The point-like particles are classified by choosing
$X$ to be $\bR^{1,3}\times S^{2,3}$, where
$\bR^{1,3}=\{(z,x^1,x^2,x^3)\}$ and considering stable
D-branes whose world-volumes are of the form
$\bR^{0,1}\times X'$,
where $\bR^{0,1}$ is the time direction and $X'$ is
a compact submanifold of $X$. The probe D8-branes
are extended along $\bR^{0,1}\times X$.
For this configuration, we have $n=\Dint=4$ and
the classification of the particles is given by
\begin{align}
&KH(\bR^{1,3}\times S^{2,3})=
KH(\bR^{1,3}) \oplus KH(\bR^{3,5})=0
\ ,~~~
(\mbox{for O6$^-$}) \nn\\
&KR(\bR^{1,3}\times S^{2,3})=
KR(\bR^{1,3}) \oplus KR(\bR^{3,5})=\bZ_2\oplus\bZ_2
\ .~~~
(\mbox{for O6$^+$})
\end{align}
Here, $KH(\bR^{1,3})$ and $KR(\bR^{1,3})$ correspond
to the D4-\AD4 pairs wrapped on the $S^4$, which are
interpreted as the baryons.
The elements of
$KH(\bR^{3,5})$ and $KR(\bR^{3,5})$ correspond
to the D0-\AD0 pairs, which do not correspond
to bound states of quarks and/or gluons
in QCD.\footnote{
The D0-brane also exists in the case without
orientifold plane. Since it has a net charge for RR 1-form
potential, it cannot be considered as a bound state
made by quarks and/or gluons.
}

The string-like objects extended along $(x^0,x^1)$
are classified by choosing
$X$ to be $\bR^{1,2}\times S^{2,3}$, where
$\bR^{1,2}=\{(z,x^2,x^3)\}$. They are classified by
\begin{align}
&KH(\bR^{1,2}\times S^{2,3})=
KH(\bR^{1,2}) \oplus KH(\bR^{3,4})=0
\ ,~~~
(\mbox{for O6$^-$}) \nn\\
&KR(\bR^{1,2}\times S^{2,3})=
KR(\bR^{1,2}) \oplus KR(\bR^{3,4})=\bZ_2\oplus\bZ_2
\ .~~~
(\mbox{for O6$^+$})
\end{align}
Here, $KH(\bR^{1,2})$ and $KR(\bR^{1,2})$ correspond
to the non-BPS D5-branes wrapped on the $S^4$, that are
interpreted as the flux tube in QCD.
The elements of
$KH(\bR^{3,4})$ and $KR(\bR^{3,4})$ correspond
to the non-BPS D1-branes, which do not have any
counterparts in QCD.

For completeness, let us sketch the derivation of (\ref{KKK}).
{}From (\ref{KRpq}), it is sufficient to show the $p=q=0$ case in
(\ref{KKK}). We use the fact that there is an exact sequence
\begin{align}
KR^{-n-1}(X)\ra KR^{-n-1}(Y)\ra KR^{-n}(X,Y)\ra KR^{-n}(X)\ra
 KR^{-n}(Y)\ ,
\label{es}
\end{align}
where $Y\subset X$, and $(X,Y)$ is a pair of compact spaces with involution.
It follows from this exact sequence that
\begin{eqnarray}
 KR^{-n}(X)=KR^{-n}(X,Y)\oplus KR^{-n}(Y)\ ,
\label{retract}
\end{eqnarray}
when $Y$ is a retract of $X$, i.e.,
there is a continuous map $r: X\ra Y$ such that
$r|_Y = 1_{Y}$.
Applying (\ref{retract}) for $X=S^{r,s}$ and $Y=pt$, we obtain
\begin{eqnarray}
 KR^{-n}(S^{r,s})=KR^{-n}(S^{r,s},pt)\oplus KR^{-n}(pt)\ .
\label{retract2}
\end{eqnarray}
Here, $Y=pt$ is chosen to be a fixed point of the involution
in $X=S^{r,s}$, which exists for $s\ge 1$.

Again, applying (\ref{retract}) for $X=B^{r,s}$ (a unit ball in $\bR^{r,s}$)
and $Y=S^{r,s}$, and using the facts
 $KR^{-n}(B^{r,s})=KR^{-n}(pt)$ and
 $KR^{-n}(B^{r,s},S^{r,s})=KR^{-n}(\bR^{r,s})=KR^{-n+r-s}(pt)$, we can show
\begin{eqnarray}
KR^{-n}(S^{r,s},pt)=KR^{-n+r-s+1}(pt)\ .
\label{KR2}
\end{eqnarray}
{}From (\ref{retract2}) and (\ref{KR2}), we obtain (\ref{KKK}).

\end{document}